\documentclass[aps,prd,twocolumn,nofootinbib]{revtex4-2}
\usepackage{epsfig}
\usepackage{graphicx}
\usepackage{dcolumn}
\usepackage{amssymb,amsmath}
\usepackage{mathrsfs}
\usepackage{epsfig,bm}
\usepackage{bm}
\begin{document}

\title{Parity Nonconservation in Rb and Sr$^+$ due to Low-Mass Vector Boson}

\author{V. A. Dzuba}\email{v.dzuba@unsw.edu.au}
\author{V. V. Flambaum}\email{v.flambaum@unsw.edu.au}
\author{G. K. Vong}\email{g.vong@unsw.edu.au}

\affiliation{School of Physics, University of New South Wales, Sydney 2052, Australia}

\begin{abstract}

We calculate the parity non-conserving (PNC) electric-dipole ($E1$) transition amplitudes for the $5s - 6s$ and $5s - 4d_{3/2}$ transitions in Rb and Sr$^+$. Our results include both the nuclear-spin-independent and nuclear-spin-dependent contributions, with particular emphasis on the potential effects of a hypothetical additional $Z'$-boson. We highlight possible  advantages of using light atoms in searches for such new interaction.
The ratio of the contribution of a low mass  $Z'$-boson to the contribution of the Standard  model $Z$ -boson to PNC effects increases rapidly (faster than $1/Z^2$) with decreasing nuclear charge $Z$. Another advantage is that  theoretical interpretations of experiments in lighter systems may be carried out with a higher accuracy than that in Cs, Ba$^+$, Fr and Ra$^+$.

\end{abstract}

\maketitle

\section{Introduction}

The Standard Model (SM) of particle physics is the most successful framework for describing elementary particles and their interactions. However, it does not account for several observed phenomena, most notably dark matter. This strongly suggests the existence of new, feebly interacting particles beyond the SM.

An additional light  $Z'$-boson is among the proposed candidates for dark matter~\cite{Nelson:2011sf} or serve as a mediator between dark matter and SM particles~\cite{Okada:2018ktp}.
Similar to the Standard Model $Z$-boson, it would  contribute to atomic parity non-conserving (PNC) effects. Therefore, precision studies of PNC in atoms can place constraints on the properties of this extra $Z'$-boson. This is achieved by comparing measured PNC amplitudes with Standard Model predictions and attributing any observed deviation to the contribution from the additional $Z$-boson.


First PNC experiments with Bi, Pb, and  Tl  atoms have provided important results for the establishing of the  Standard model (see e.g. reviews \cite{Khriplovich,Ginges,DK04,DF-PNC12}).  However, the accuracy of atomic calculations for these complex atoms  is significantly lower than the experimental accuracy. 
 The highest accuracy of the relativistic many-body calculations ($0.5$\% in Cs) is achieved  in atoms with one valence electrons, where all-orders summation of dominating chains of the correlation diagrams is possible \cite{DzuFlaSus89,Blundell1990,DzuFlaGin02,DereviankoCsPNC,CsPNC12,Sahoo2021}.
The best  experimental  accuracy 0.35\% has been obtained  in the measurement of the $6s$-$7s$ PNC amplitude in cesium ~\cite{Wood}.  The sub-percent agreement between experiment and theory in Cs has enabled stringent constraints  on possible new interactions beyond the Standard Model (see e.g. ~\cite{DzuFlaGin02,DereviankoCsPNC,CsPNC12,Sahoo2021}).

Further improvement in  the  search for new physics in atomic PNC experiments is limited by  the accuracy of atomic calculations. 
Rubidium was proposed as a promising candidate for PNC studies in Ref.~\cite{RbPNC12}. Its electronic structure is similar to that of cesium (but with a smaller number of core electrons), allowing for theoretical calculations with a comparable or even better  accuracy. 

Moreover, its lower nuclear charge $Z$ leads to significantly reduced contributions from several corrections, such higher order terms in quantum electrodynamics (QED)   and relativistic Breit-type   corrections, as well as from neutron skin correction (difference between the neutron and proton distributions in nuclei, which is very hard to measure  due to electric neutrality of neutrons). These corrections  rapidly increase with nuclear charge and may be significant  in heavier atoms Cs, Ba$^+$, Fr and Ra$^+$ with a similar electronic structure.  As a result, a  higher overall theoretical accuracy can be achieved for Rb.

 Low sensitivity to neutron skin  in light elements is especially important for the isotope chain method suggested in Ref. \cite{DzuFlaKhr}.  In the ratio of  PNC effects in different isotopes an error of atomic calculations cancels out. The dominating theoretical error comes  from the uncertainties in the neutron distributions. Sensitivity to the neutron skin, investigated in Refs. \cite{Fortson,Brown,Viatkina}, is proportional to $Z^2 \alpha^2$ and is suppressed in light elements. First experimental implementation in Yb isotope chain was published in Refs. \cite{Antypas2018,Antypas2019}.  Strontium has  4 stable isotopes, $^{84}$Sr, $^{86}$Sr, $^{87}$Sr, and $^{88}$Sr,  and significantly weaker sensitivity to the neutron skin than Yb. Combination of single-isotope and isotope chain methods allows one to measure separately both proton and neutron weak charges. 

In the present work, we extend our earlier study of PNC in Rb to include the contribution of an additional vector $Z'$-boson  with arbitrary mass $m_{Z'}$. This analysis closely follows the approach previously applied to Cs, Ba$^+$, Yb, Tl, Fr, and Ra$^+$ in Ref.~\cite{DFS17}. 

Our primary goal is to demonstrate that the ratio  of the $Z'$-boson contribution to the Standard model contribution  is enhanced in lighter atomic systems. We therefore also consider the Sr$^+$ ion, which has an electronic structure analogous to that of Rb, and calculate the $5s - 6s$ and $5s - 4d_{3/2}$ PNC amplitudes in both systems. 

The present calculations are performed with  accuracy about 1\% for $5s - 6s$ and a few per cent for   $5s - 4d_{3/2}$ transitions for the Standard model results. We do not aim for  a higher accuracy here,  since the  main purpose  of the present paper is  to determine the relative contribution of new physics effects. 
All major contributions, such as electron correlation effects,
 core polarisation, and the Breit interaction are included. At this stage, we neglect  smaller corrections, such as QED effects, structure radiation, and the neutron skin contribution.

\section{PNC amplitudes}

The Hamiltonian describing the parity-nonconserving electron-nuclear interaction can be written as a sum of the spin-independent (SI) and spin-dependent (SD) parts:
\begin{equation}\label{e:HPNC}
    \hat{H}_\text{PNC} = \hat{H}_\text{SI} + \hat{H}_\text{SD}  = \frac{G_F}{\sqrt{2}} \left( - \frac{Q_W}{2} \gamma_5 + \frac{\varkappa}{I} \boldsymbol{\alpha} \boldsymbol{I} \right) \rho(\boldsymbol{r}),
\end{equation}
where $G_F$
is the Fermi weak interaction constant, $Q_W$ is the nuclear weak charge, $\boldsymbol{\alpha} 
$ and $\gamma_5$ are the Dirac matrices, $\boldsymbol{I}$ is the nuclear spin, and $\rho(\boldsymbol{r})$ is the nuclear density normalised to 1. The dimensionless constant $\varkappa$ determines the strength of the SD PNC interaction and is to be found from measurements. The weak nuclear charge \( Q_W \) is given by~\cite{SM}
\begin{equation}
    Q_W \approx -0.9877\,N + 0.0716\,Z. 
\end{equation}
Here \(N\) is the number of neutrons, and \(Z\) is the number of protons.

For the exchange of a vector boson $Z'$
between electrons and nucleons, we adopt the following couplings:
\begin{equation}
    L_\text{int} = Z_{\mu}' \sum_{f=e,p,n} \bar{f} \gamma^\mu \left( g_f^V + \gamma_5 g_f^A \right)f.
\end{equation}
The resulting P-violating interaction between two fermions is described by a Yukawa-type potential:
\begin{equation}
    V_{12}(r) = \frac{g_1^A g_2^V}{4\pi} \frac{e^{-m_{Z'} r}}{r} \gamma_5,
    \label{e:V12}
\end{equation}
where $r$ is the distance between two fermions, the $\gamma$ matrix corresponds to fermion 1 (electron) , and we have treated fermion 2 (proton or neutron) non-relativistically. We introduce the shorthand notation $g_v^N = (N g_n^V + Z g_p^V)/A$, where $A= Z + N $ is the nucleon number.

The PNC amplitude 
for the atomic transition $a \to b$ is given by
\begin{equation}\label{e:amp}
    E_{\text{PNC}}^{a \to b} = \sum_n \left[ \frac{\langle a | \hat W | n  \rangle \langle n | \mathbf{D} | b  \rangle}{E_a - E_n} + \frac{\langle a | \mathbf{D} | n  \rangle \langle n | \hat W | b  \rangle}{E_b - E_n} \right],
\end{equation}
where $\mathbf{D}$ is the electric dipole operator and  weak interaction Hamiltonian $\hat W$ is given by either (\ref{e:HPNC}) or (\ref{e:V12}).

The methods of calculating wave functions and matrix elements are presented in the appendix.

\section{Results and discussion}

\begin{table}
 \caption{\label{t:PNC} 
Nuclear spin independent PNC amplitudes in Rb and Sr$^+$ induced by first term in (\ref{e:HPNC}).}
\begin{ruledtabular}
\begin{tabular}{l l l l c c}
\multicolumn{1}{c}{Atom}&
\multicolumn{1}{c}{$Q_W$}&
\multicolumn{1}{c}{$I$}&
\multicolumn{1}{c}{Transition}&
\multicolumn{1}{c}{$A_{\rm PNC}$}&
\multicolumn{1}{c}{$A_{\rm PNC}$}\\

&&&&\multicolumn{1}{c}{$10^{-14} iea_B Q_W$}&
\multicolumn{1}{c}{$10^{-12} iea_B$}\\
\hline 
$^{85}$Rb     & -44.76 & 5/2 & $5s - 6s$       & 2.897 & -1.30 \\
              &        &     & $5s - 4d_{3/2}$ & 9.030 & -4.04 \\
$^{87}$Rb     & -46.74 & 3/2 & $5s - 6s$       & 2.897 & -1.35 \\
              &        &     & $5s - 4d_{3/2}$ & 9.030 & -4.22 \\
$^{87}$Sr$^+$ & -45.68 & 9/2 & $5s - 6s$       & 2.105 & -0.96 \\
              &        &     & $5s - 4d_{3/2}$ & 5.606 & -2.56 \\
\end{tabular}
\end{ruledtabular}
\end{table}

The calculated nuclear-spin-independent PNC amplitudes in Rb and Sr$^+$ are presented in Table~\ref{t:PNC}. Our results for the $5s - 6s$ amplitude in Rb are in good agreement with previous calculations~\cite{DzuFlaSilSus87a,RbPNC12}, while the other amplitudes have not been studied previously. We note that the PNC amplitudes for the $5s - 4d_{3/2}$ transitions are about three times larger than those for the $5s - 6s$ transitions. This behaviour is consistent that observed in Cs, Ba$^+$~\cite{DzuFlaGin01}, and other atoms and ions with similar electronic structure.

The $s - d$ PNC transitions has  an important property  ~\cite{DzuFlaGin01}: the sum in Eq.~(\ref{e:amp}) is strongly dominated by a single term (the $n=5p_{1/2}$ intermediate state in the first term of Eq.~(\ref{e:amp}) for both Rb and Sr$^+$). As a result, precise experimental data for the $\langle 5p_{1/2}|\mathbf{D}|5d_{3/2}\rangle$ E1 transition matrix element can be used to improve the accuracy of the theoretical results.

\begin{table}
  \caption{\label{t:PNCAM} 
PNC amplitudes ($z$-components) for the $|5s,F_1 \rangle - |6s,F_2 \rangle$ and  $|5s,F_1 \rangle - |4d_{3/2},F_2 \rangle$ transitions 
in $^{87}$Sr$^+$. 
Units are $10^{-11} iea_B$.}
\begin{ruledtabular}
\begin{tabular}   {ll cc r}
\multicolumn{1}{c}{Isotope}&
\multicolumn{1}{c}{Transition}&
\multicolumn{1}{c}{$F_1$}&
\multicolumn{1}{c}{$F_2$}&
\multicolumn{1}{c}{PNC amplitude}\\
\hline 
$^{87}$Sr$^+$ & $5s - 6s$      &  4 &  4 &  $0.0770[1+0.0447\varkappa$] \\
              &                &  4 &  5 & $-0.0577[1-0.0868\varkappa$] \\
              &                &  5 &  4 & $-0.0577[1+0.0950\varkappa$] \\
              &                &  5 &  5 & $-0.0963[1-0.0366\varkappa$] \\
&&&& \\	       		             	  
$^{87}$Sr$^+$ & $5s-4d_{3/2}$  &  4 &  3 &  $0.1479[1+0.0533\varkappa$] \\
              &                &  4 &  4 &  $0.2569[1+0.0533\varkappa$] \\
              &                &  4 &  5 & $-0.0948[1+0.0533\varkappa$] \\
              &                &  5 &  4 &  $0.0700[1-0.0436\varkappa$] \\
              &                &  5 &  5 &  $0.2371[1-0.0436\varkappa$] \\
              &                &  5 &  6 & $-0.1369[1-0.0436\varkappa$] \\
\end{tabular}
\end{ruledtabular}
\end{table}

Table~\ref{t:PNCAM} presents the combined nuclear-spin-independent and nuclear-spin-dependent PNC amplitudes for different hyperfine transitions of the $5s - 6s$ and $5s - 4d_{3/2}$ transitions in Sr$^+$. Corresponding results for Rb can be found in our earlier works~\cite{RbPNC12,DFS17}. These data enable the extraction of the parameter $\varkappa$ (which includes, in particular, the contributions of the nuclear anapole moment and $Z$ boson exchange) from measurements of the PNC amplitudes.

\begin{table}
  \caption{\label{t:mz} 
PNC amplitudes induced by interaction (\ref{e:V12}) for various vector-boson masses.
The presented values for the atomic PNC amplitudes are in terms of the parameter 
$-Q_W/N = (2\sqrt{2}Ag_e^A g_N^V )/(NG_F m_{Z'}^2 )$ 
and in the units  $iea_B$ (see also \cite{DFS17}).
}
\begin{ruledtabular}
\begin{tabular}   {l rrrr}
\multicolumn{1}{c}{$m_{Z'}$}&
\multicolumn{2}{c}{$^{87}$Rb}&
\multicolumn{2}{c}{$^{87}$Sr$^+$}\\
\multicolumn{1}{c}{(eV)}&
\multicolumn{1}{c}{$5s-6s$}&
\multicolumn{1}{c}{$5s-4d_{3/2}$}&
\multicolumn{1}{c}{$5s-6s$}&
\multicolumn{1}{c}{$5s-4d_{3/2}$}\\
\hline %
$10^9$ &  $-1.3 \times 10^{-12}$ &   $-4.2 \times 10^{-12}$ &   $-9.8 \times 10^{-13}$ &   $-2.8 \times 10^{-12}$ \\
$10^8$ &  $-1.3 \times 10^{-12}$ &   $-4.2 \times 10^{-12}$ &   $-9.6 \times 10^{-13}$ &   $-2.8 \times 10^{-12}$ \\
$10^7$ &  $-1.1 \times 10^{-12}$ &   $-3.6 \times 10^{-12}$ &   $-8.3 \times 10^{-13}$ &   $-2.4 \times 10^{-12}$ \\
$10^6$ &  $-7.2 \times 10^{-13}$ &   $-2.3 \times 10^{-12}$ &   $-5.1 \times 10^{-13}$ &   $-1.5 \times 10^{-12}$ \\
$10^5$ &  $-9.6 \times 10^{-14}$ &   $-3.4 \times 10^{-13}$ &   $-6.7 \times 10^{-14}$ &   $-7.8 \times 10^{-14}$ \\
$10^4$ &  $-1.6 \times 10^{-15}$ &   $-7.0 \times 10^{-15}$ &   $-1.1 \times 10^{-15}$ &   $ 3.5 \times 10^{-15}$ \\
$10^3$ &  $-8.2 \times 10^{-18}$ &   $-2.4 \times 10^{-17}$ &   $-6.9 \times 10^{-18}$ &   $-2.4 \times 10^{-17}$ \\
$10^2$ &  $-4.4 \times 10^{-20}$ &   $-1.1 \times 10^{-19}$ &   $-5.8 \times 10^{-20}$ &   $-3.4 \times 10^{-19}$ \\
$10$   &  $-4.3 \times 10^{-22}$ &   $-1.1 \times 10^{-21}$ &   $-5.8 \times 10^{-22}$ &   $-3.4 \times 10^{-21}$ \\
\end{tabular}
\end{ruledtabular}
\end{table}

\begin{figure}
\epsfig{figure=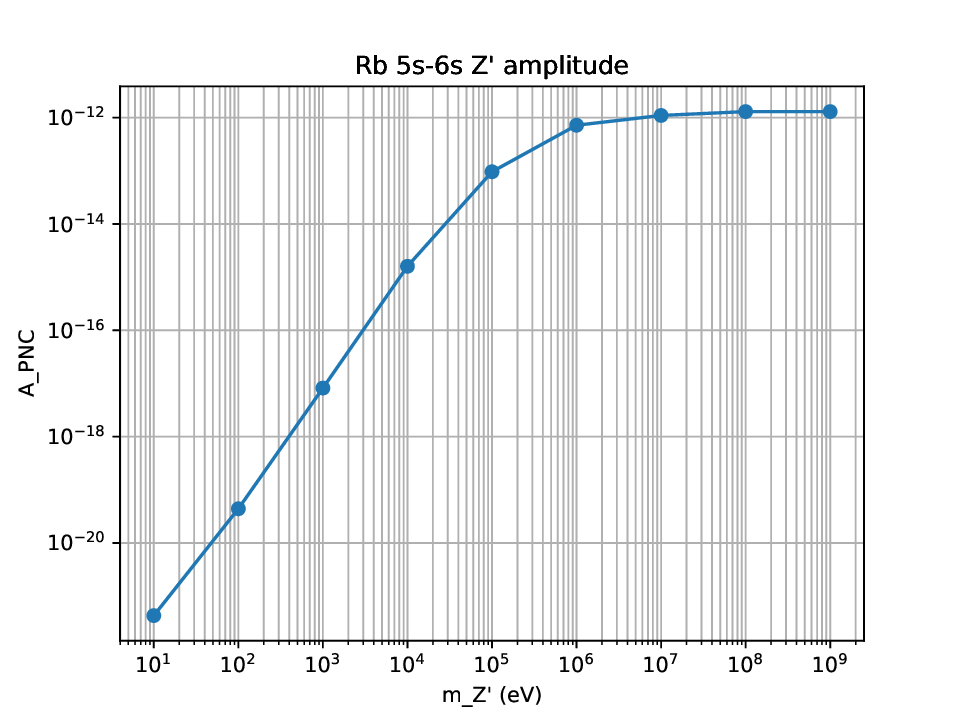,scale=0.5}
\caption{PNC amplitude cause by $Z"$ boson as a function of its mass, see Table~\ref{t:mz}.}
\label{f:rb}
\end{figure}

Table~\ref{t:mz} shows the contribution of a $Z'$ boson to the PNC amplitudes in Rb and Sr$^+$ as a function of its mass. The units are chosen such that, in the limit $m_{Z'} \rightarrow \infty$, the $Z'$ contribution reduces to that of the Standard Model $Z$ boson. Therefore, the values in the first row of Table~\ref{t:mz} are close to  the corresponding results in the last column of Table~\ref{t:PNC}. The data for the $5s - 6s$ transition in Rb are presented also on Fig.~\ref{f:rp}.

The last rows of Table~\ref{t:mz} represent the low-mass limit, where the range of the Yukawa potential exceeds the size of the atom. In this regime, the $Z'$ exchange amplitude does not depend on $m_{Z'}$. Because of the choice of units (which are proportional to $1/m_{Z'}^2$),  the last numerical values in   Table~\ref{t:mz}  scales as $m_{Z'}^2$. 

The results in Table~\ref{t:mz} can be used to find values of  the strength of the new interaction mediated by the $Z'$ boson by attributing any possible difference between measured PNC amplitudes and Standard Model predictions to the presence of the $Z'$ contribution.

The  $Z'$ interaction constants  can be found  using expressions
\begin{equation}\label{e:x}
\frac{g^A_e g^V_N}{m_{Z'}^2} = \frac{G_F}{2\sqrt{2}}\frac{N}{A}\frac{A_{\rm PNC}}{A_{\mathrm{PNC}}^{Z'}}\varepsilon, \ \ m_{Z'} \gg Z\alpha m_e,
\end{equation}
and
\begin{equation}\label{e:y}
g^A_e g^V_N = \frac{G_F}{2\sqrt{2}}\frac{N}{A}\frac{A_{\rm PNC}}{A_{\mathrm{PNC}}^{Z'}} m_{Z'}^2 \varepsilon, \ \ m_{Z'} \ll 1/R_{\rm atom}.
\end{equation}

In both expressions, $N$ is the neutron number, $A$ is the mass number, $A_{\rm PNC}$ is  the tabulated numerical value of the spin-independent PNC amplitude expressed in units of $i e a_B$,
see Table  \ref{t:PNC}; $A_{\mathrm{PNC}}^{Z'}$ is the tabulated numerical value of the $Z'$-boson contribution to the PNC amplitude.
 The values of $A_{\mathrm{PNC}}^{Z'}$ used in Eq.~(\ref{e:x}) should be taken from the first row of Table~\ref{t:mz}, while those for Eq.~(\ref{e:y}) correspond to the last row of the table.  Parameter $\varepsilon$ denotes the relative deviation of the experimental PNC amplitude from the Standard Model prediction. 

For  illustration, we assume $\varepsilon = 0.001$ for both Rb and Sr$^+$. 
This optimistic choice corresponds to roughly an order-of-magnitude improvement in accuracy relative to cesium
~\cite{Wood,DzuFlaGin02,DereviankoCsPNC,CsPNC12}.   Such an assumption may be justified by the expectation that several small contributions to the PNC amplitude - including higher-order Breit-type relativistic effects, higher-order QED corrections, and the neutron-skin contribution - are substantially suppressed in lighter atoms.

\begin{figure}
\epsfig{figure=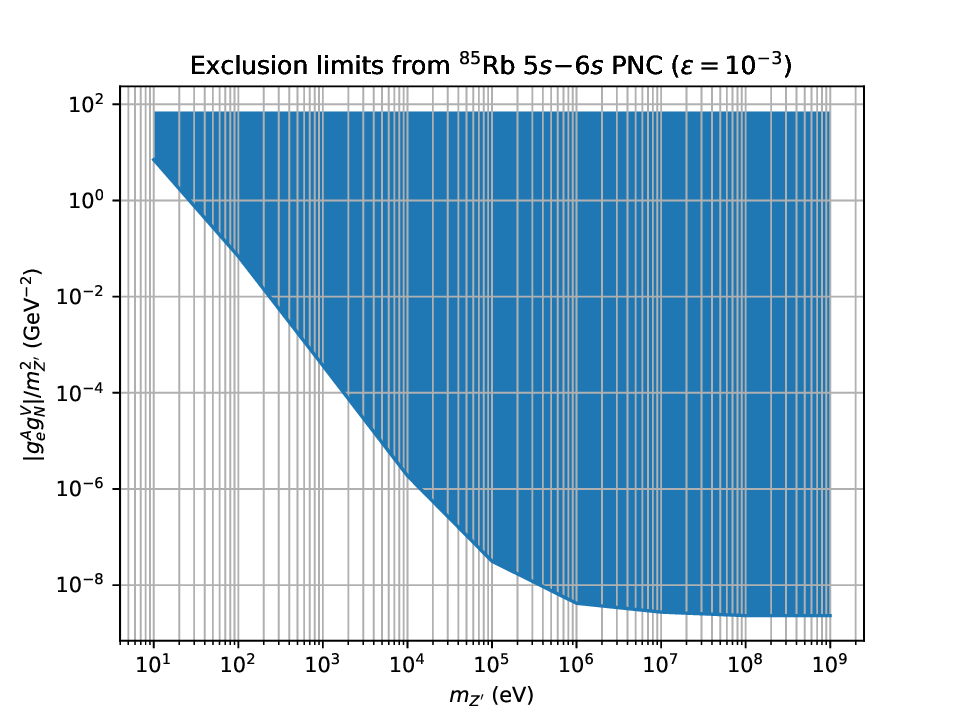,scale=0.5}
\caption{Hypothetical exclusion plot for the $g_e^Ag_N^V/m_{Z'}^2$ combination of parameters obtained from the data in Table~\ref{t:mz} under assumption that accuracy of the measurements is $10^{-3}$.}
\label{f:m2}
\end{figure}

\begin{figure}
\epsfig{figure=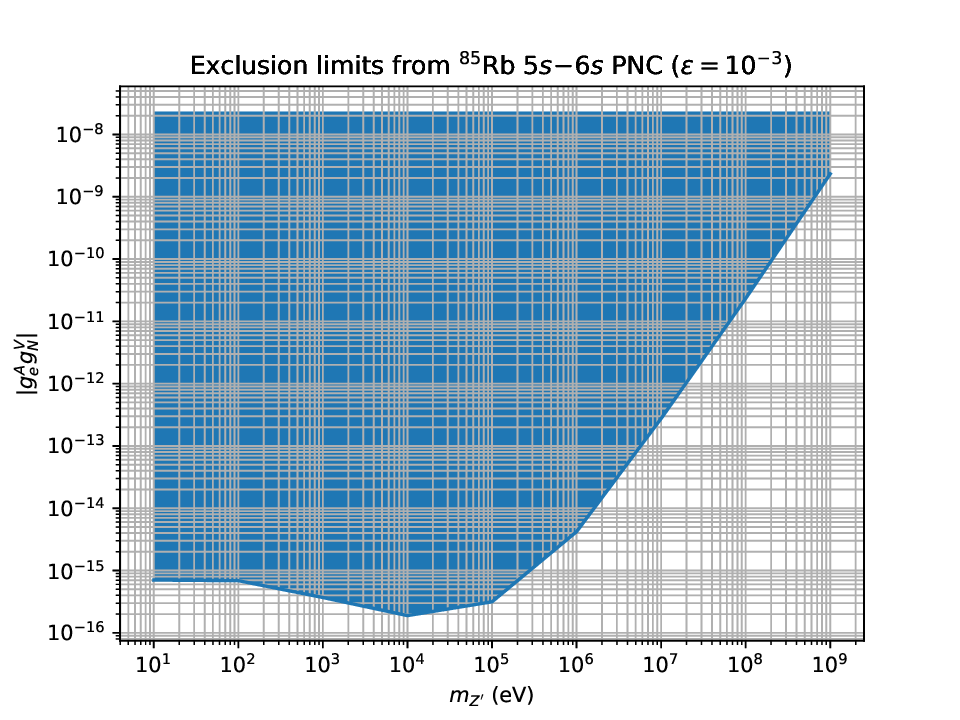,scale=0.5}
\caption{Exclusion plot for the $g_e^Ag_N^V$ combination of parameters.}
\label{f:g}
\end{figure}

The estimates obtained using Eqs.~(\ref{e:x}) and (\ref{e:y}), together with the data from Tables~\ref{t:PNC} and \ref{t:mz}, are summarized in Table~\ref{t:limits}. For comparison, we also list the limits extracted from Cs data in Ref.~\cite{DFS17}. For large $m_{Z'}$, we observe roughly an order-of-magnitude potential improvement in the constraints, if assume $\varepsilon = 0.001$.  This directly reflects our assumption of a ten-fold increase in the experimental and theoretical accuracy relative to Cs.
The hypothetical exclusion plots for the parameters of the interaction involving $Z'$ boson which might be obtained from the measurements of the PNC in the $5s - 6s$ transition in Rb are presented on Figs.~\ref{f:m2} and \ref{f:g}.

The behavior at small $m_{Z'}$  is different. The Standard Model weak interaction matrix element includes an additional factor $Z^2 R$ (where $R$ is the relativistic enhancement factor; $R \sim  10$ in heavy atoms). This factor cancels out in Eq.~(\ref{e:x}) in the large-mass limit but becomes important in the small-mass regime described by Eq.~(\ref{e:y}), where it suppresses the relative sensitivity  to $Z'$ in heavier atoms. As a result, a possible improvement in sensitivity to a light $Z'$ boson reaches a factor of about 40 in Rb and Sr$^+$ compared with Cs,  assuming $\varepsilon = 0.001$.  Using results of Table~\ref{t:limits}, one can estimate the sensitivity to $Z'$ for any other  $\varepsilon$. 

\begin{table}
  \caption{\label{t:limits} Possible limits on the combinations of parameters $g^A_e g^V_N/m_{Z'}^2$
 for $m_Z'  \gg Z\alpha m_e$ and $g^A_e g^V_N$ for $m_{Z'}  \ll R_{\rm atom}$,
 from the consideration of vector-mediated P-violating interactions in atoms, assuming the relative accuracy  $\varepsilon = 0.001$. The data for Cs are taken from \cite{DFS17} and included for comparison.}
\begin{ruledtabular}
\begin{tabular}   {llll}
\multicolumn{1}{c}{Atom}&
\multicolumn{1}{c}{Transition}&
\multicolumn{1}{c}{$|g_e^Ag_N^V|/m_{Z'}^2$}&
\multicolumn{1}{c}{$|g_e^Ag_N^V|$}\\
&& \multicolumn{1}{c}{[GeV$^{-2}$]}&\\
\hline %
$^{133}$Cs & $6s - 7s$       & $3.9 \times 10^{-8}$ & $3.1 \times 10^{-14}$ \\
 $^{85}$Rb & $5s - 6s$       & $2.5 \times 10^{-9}$ & $7.5 \times 10^{-16}$ \\
 $^{85}$Rb & $5s - 4d_{3/2}$ & $2.4 \times 10^{-9}$ & $9.2 \times 10^{-16}$ \\
 $^{87}$Sr$^+$ & $5s - 6s$       & $2.4 \times 10^{-9}$ & $4.1 \times 10^{-16}$ \\
 $^{87}$Sr$^+$ & $5s - 4d_{3/2}$ & $2.3 \times 10^{-9}$ & $1.9 \times 10^{-16}$ \\
\end{tabular}
\end{ruledtabular}
\end{table}

\section{Conclusion}

In this work, we have performed calculations of parity non-conservation (PNC) amplitudes for selected transitions in neutral Rb and singly ionised Sr$^+$. The dependence of the PNC amplitudes on the mass of a hypothetical vector boson has been investigated over a wide range, covering both the contact-interaction regime and the finite-range limit.

The results show that, for large boson masses, the interaction reduces to the standard contact form, while for lighter bosons the finite interaction range leads to a suppression of the PNC amplitudes. This behavior is reflected in the calculated matrix elements and provides the basis for deriving constraints on the product of coupling constants as a function of the boson mass. Among the considered transitions, those with larger electronic overlap in the nuclear region exhibit enhanced sensitivity to new interactions.

The calculated amplitudes can be combined with experimental data to set exclusion limits on new vector bosons over a broad mass range. Atomic PNC measurements in Rb and Sr$^+$ thus provide a sensitive probe of physics beyond the Standard Model, complementary to high-energy searches. Further improvements in both experimental accuracy and the theoretical treatment of correlations will enhance the sensitivity of these tests. The projected sensitivity represents an improvement by approximately a factor of 40 compared to existing constraints derived from PNC measurements in Cs.

\begin{acknowledgments}

This work was supported by the Australian Research Council Grant No. DP230101058.
\end{acknowledgments}

\appendix 
\section{Method of calculation}

In this work we follow the methods developed for highly accurate calculations of PNC in Cs~\cite{DzuFlaSus89}.
The calculations start from the relativistic Hartree-Fock (RHF) calculations for the closed-shell core.
Then, the all-order correlation potential $\hat \Sigma$ is calculated using the Feynman diagram technique~\cite{DzuFlaSus89}.
The states of the valence electron, which include the effect of correlations (the so-called Brueckner orbitals (BO))
are found by solving the Hartree-Fock-like equations with an extra operator $\hat \Sigma$: 
\begin{equation}
 (\hat H_0 +\hat \Sigma - \epsilon_v)\psi_v^{(\rm BO)}=0.
\label{eq:BO}
\end{equation}
Here $\hat H_0$ is the RHF Hamiltonian, which includes the Breit interaction~\cite{DzuEtAl01a}, index
$v$ numerate valence states. The BO $\psi_v^{(\rm BO)}$ and energy
$\epsilon_v$ include dominating higher-order correlations. 
 
The parity non-conserving weak interaction as well as the electric
dipole interaction of the atom with laser fiels are included in the
framework of the  random-phase approximation (RPA, see, e.g. \cite{DzuFlaSilSus87}).

In the RPA method, a single-electron wave function in external weak and
$E1$ electric dipole fields is
\begin{equation}
\psi = \psi_0 + \delta\psi +X e^{-i\omega t}+Y e^{i\omega t} + \delta
Xe^{-i\omega t} + \delta Ye^{i\omega t},
\end{equation}
where $\psi_0$ is the unperturbed state, $\delta\psi$ is the
correction due to the weak interaction acting alone, $X$ and $Y$ are
corrections due to the photon field acting alone, $\delta X$ and
$\delta Y$ are corrections due to both fields acting simultaneously,
and $\omega$ is the frequency of the PNC transition. The corrections
are found by solving  the system of RPA equations self-consistently 
for all core states
\begin{eqnarray}
&&(\hat H_0 - \epsilon_c)\delta\psi_c = - (\hat H_W + \delta \hat
V_W)\psi_{0c}, \nonumber \\ 
&&(\hat H_0 - \epsilon_c-\omega)X_c = - (\hat H_{E1} + \delta \hat
V_{E1})\psi_{0c}, \nonumber \\ 
&&(\hat H_0 - \epsilon_c+\omega)Y_c = - (\hat H_{E1}^{\dagger} + \delta \hat
V_{E1}^{\dagger})\psi_{0c}, \label{eq:RPA} \\ 
&&(\hat H_0 - \epsilon_c-\omega)\delta X_c = -\delta\hat V_{E1}\delta\psi_c -
\delta\hat V_WX_c - \delta\hat V_{E1W}\psi_{0c}, \nonumber \\ 
&&(\hat H_0 - \epsilon_c+\omega)\delta Y_c = -\delta\hat
V_{E1}^{\dagger}\delta\psi_c -
\delta\hat V_WY_c - \delta\hat V_{E1W}^{\dagger}\psi_{0c}, \nonumber 
\end{eqnarray}
where index $c$ numerates core states, $\hat H_W$ is the Hamiltonian of the weak interaction (first or second term of (\ref{e:HPNC}))
$\delta \hat V_W$ and
$\delta \hat V_{E1}$ are corrections to the core potential due to the
weak and $E1$ interactions, respectively, and $\delta \hat V_{E1W}$ is
the correction to the core potential due to the simultaneous action of
the weak field and the electric field of the photon.

The PNC amplitude between valence states $a$ and $b$ is given by
\begin{eqnarray} \label{amp}
&& E_{\rm PNC} = \langle \psi_b^{(\rm BO)}|\hat H_{E1} + 
 \delta\hat V_{E1}|\delta\psi_a^{(\rm BO)}\rangle + \\
&& \langle \psi_b^{(\rm BO)}|\hat H_{W} + \delta\hat V_{W}|X_a^{(\rm BO)}\rangle +
\langle \psi_b^{(\rm BO)}|\delta\hat V_{E1W}|\psi_a^{(\rm BO)}\rangle. \nonumber
\end{eqnarray}
This expression includes leading correlations to all orders (via use of the BO) and core polarization by external fields,
including double core polarisation~\cite{RDF13a}.


%

\end{document}